\documentclass[12]{iopart}

\usepackage{epsfig}
\usepackage{caption}
\begin{document}

\pagestyle{empty}
\begin{center}
{\Large \bfseries
Rapidity and transverse momentum dependence of $\pi^{\scriptstyle -}\pi^{-}$
Bose-Einstein correlations measured at 20, 30, 40, 80, and 158 A$\cdot$GeV beam energy}
\end{center}
\vspace*{0.2cm}
\renewcommand{\thefootnote}{\fnsymbol{footnote}}
\noindent
S.~Kniege$^{9,*}$\footnote[0]{* parallel talk presented at Quark Matter 2004}, C.~Alt$^{9}$, T.~Anticic$^{21}$, B.~Baatar$^{8}$,D.~Barna$^{4}$,
J.~Bartke$^{6}$, 
L.~Betev$^{9,10}$, H.~Bia{\l}\-kowska$^{19}$, A.~Billmeier$^{9}$,
C.~Blume$^{9}$,  B.~Boimska$^{19}$
% ...
\begin{center}
(NA49 Collaboration)
\end{center}

\vspace{0.5cm}
\noindent
% ...
$^{4}$KFKI Research Institute for Particle and Nuclear Physics, Budapest, Hungary.\\
% ...
$^{6}$Institute of Nuclear Physics, Cracow, Poland.\\
% ...
$^{8}$Joint Institute for Nuclear Research, Dubna, Russia.\\
$^{9}$Fachbereich Physik der Universit\"{a}t, Frankfurt, Germany.\\
$^{10}$CERN, Geneva, Switzerland.\\
% ...
$^{19}$Institute for Nuclear Studies, Warsaw, Poland.\\
% ...
$^{21}$Rudjer Boskovic Institute, Zagreb, Croatia.\\

\begin{center}
\begin{parbox}[t]{12cm}{
{\small
{\bfseries Abstract:} Preliminary results on $\pi^{\scriptstyle -}\pi^{\scriptstyle -}$ Bose-Einstein correlations in 
central Pb{\footnotesize +}Pb collisions measured by the NA49 experiment are presented. 
Rapidity as well as transverse momentum dependence of the HBT- radii are shown for collisions at 
20, 30, 40, 80, and 158 A$\cdot$GeV beam energy. 
Including results from AGS and RHIC experiments only a weak energy dependence of the radii is observed.
Based on hydrodynamical models parameters like lifetime and geometrical radius 
of the source are derived from the dependence of the radii on transverse momentum.
}}
\end{parbox}
\end{center}

\setlength{\textwidth}{11cm}
\section{Introduction}

The objective of HBT analysis is to obtain information about the spatial and temporal
evolution of high energy collisions by means of two-particle Bose-Einstein
correlations. 
Two particle correlations in general lead to a
difference between the product of single-particle distributions $P(1)P(2)$ and
the corresponding two-particle distribution $P(1,2)$. The symmetrisation of the 
two particle wave function for bosons results in an enhancement of pairs with small relative momenta of the particles. 
We construct the correlation function $C_{2}(q)$ as the normalised ratio of a 
distribution of the momentum difference of true pairs {$S(q)$} and a
mixed event background distribution {$B(q)$}:
\begin{eqnarray} 
\qquad \quad C_{2}(q) & = & \frac{P(1,2)}{P(1)P(2)} = N\cdot\frac{S(q)}{B(q)}.
\end{eqnarray}
Due to space-momentum correlations in an expanding source the measured correlations do not reflect the whole extension of the source.
This enables us to study the space time evolution of the source by
analysing the correlation function in different bins
of mean transvere momentum $k_{t}$ and pair rapidity $Y$: 
\begin{eqnarray}
k_{t}=\frac{1}{2}|\vec{p}_{t,1}+\vec{p}_{t,2}| &\quad ,\quad & Y=\frac{1}{2}\log\left(\frac{E_{1}+E_{2}+p_{z,1}+p_{z,2}}{E_{1}+E_{2}-p_{z,1}-p_{z,2}}\right).\qquad
\end{eqnarray}

\begin{figure}[b]

\begin{minipage}{4.9cm}
\includegraphics[scale=0.27]{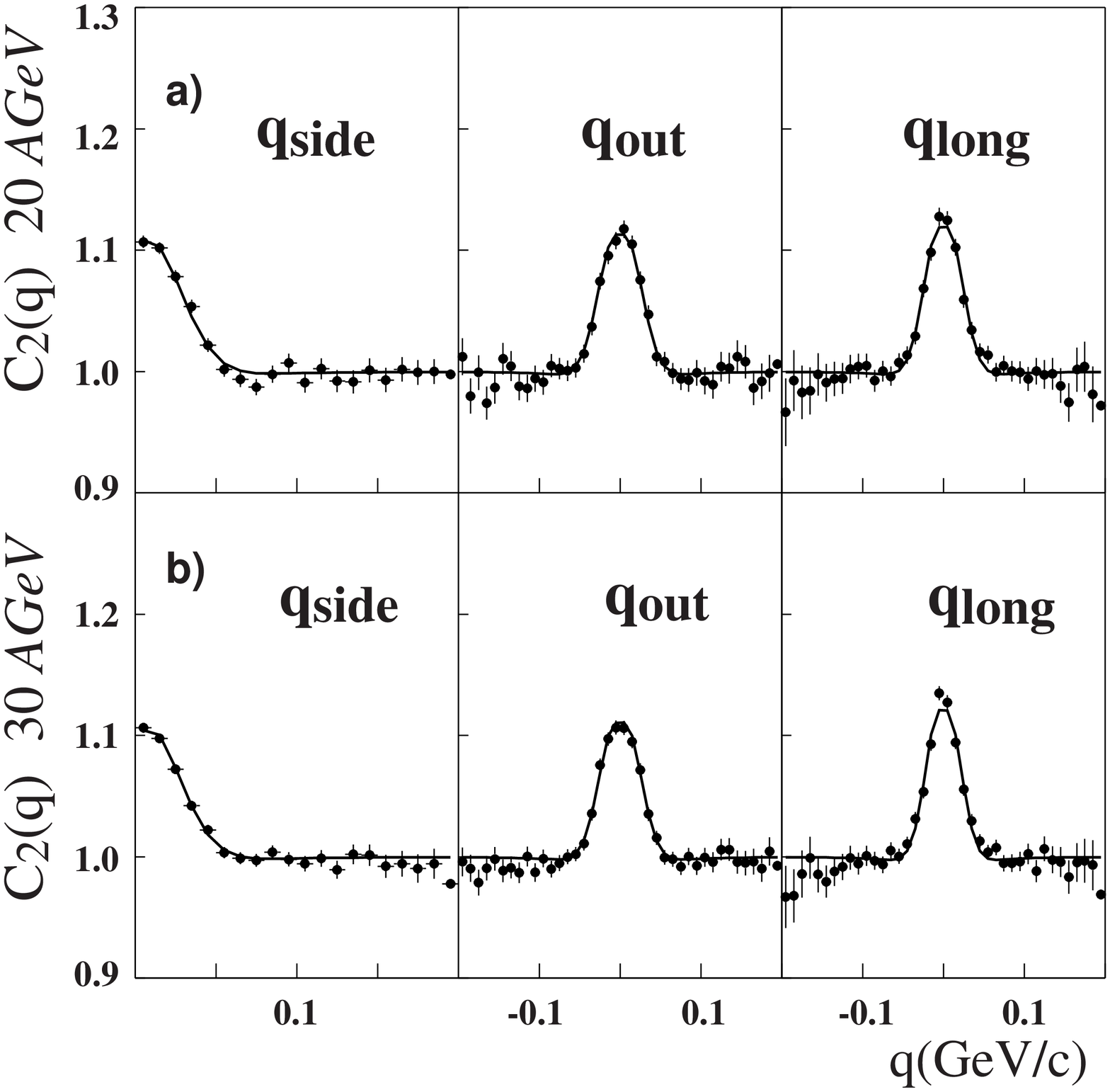}

\end{minipage}
\begin{minipage}{4,9cm}
\includegraphics[scale=0.27]{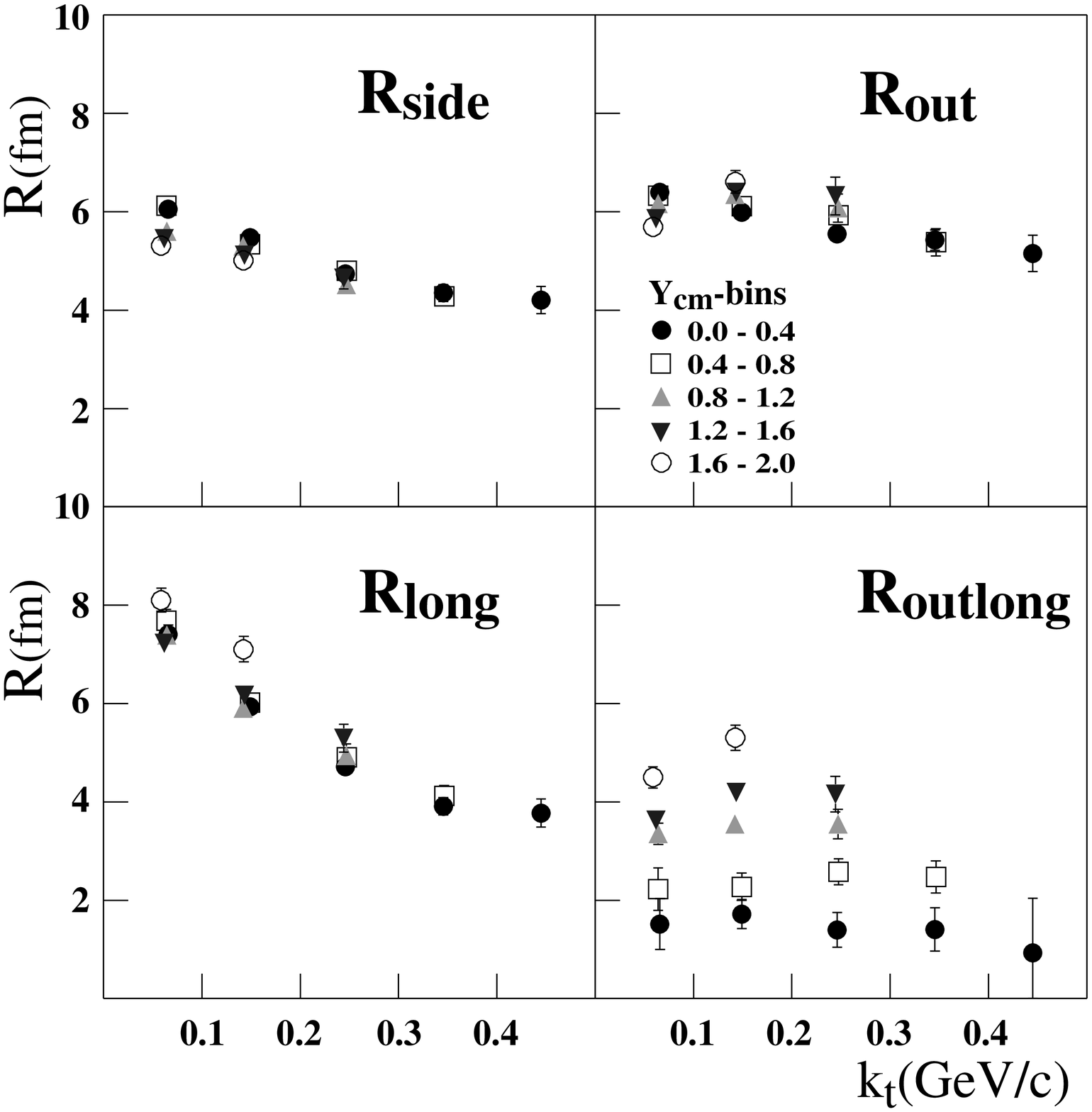}
\end{minipage} 

\begin{center}
\begin{minipage}{11.8cm}
\begin{center}
\parbox[t]{5.2cm}{
\footnotesize{{\bf Figure 1 :} Projections of the 3-dimensional correlation function and the corresponding fits of eqn.4 (solid lines) onto the components 
$q_{side}$, $q_{out}$, and $q_{long}$ for 20 (a) and 30 A$\cdot$GeV (b) in the $k_{t}$-bin (0.0-0.1) GeV/c at midrapidity. Projection range : 0.03 GeV/c.}}\hfill
\parbox[t]{5.2cm}{
\footnotesize{{\bf Figure 2 :} Rapidity and transverse momentum dependence of the HBT-radii 
$R_{side}$, $R_{out}$, $R_{long}$ and $R_{outlong}$ for 30 A$\cdot$GeV beam energy.
}}
\end{center}
\end{minipage}
\end{center}
\end{figure}

\section{Experiment and Analysis}
NA49 \cite{A.1} is a fixed target experiment located at the CERN SPS which 
comprises four large-volume Time Projection Chambers (TPC). 
A zero degree calorimeter at the downstream end of the experiment is used to trigger on the 
centrality of the collisions. 
`The data presented here correspond to the 7.2\% most central events for 20, 30, 40, and 80 A$\cdot$GeV and 
the 10\% most central events for 160 A$\cdot$GeV.
Measuring the specific energy loss dE/dx of charged particles in the
gas of the TPC's allows for particle identification with a resolution
of 3-4\%. However, due to ambiguities in particle identification by specific energy loss measurements 
in certain regions of phase space negative hadrons rather than
identified negative pions are used for the analysis. 
The limited two track resolution of the detector is taken into account by rejecting 
pairs with an average distance of particles less than 2.0 cm in the signal- as well as in the background
distribution.
Following the approach of 
Pratt and Bertsch \cite{A.2} the momentum difference is decomposed into a component parallel to the beam axis
$q_{long}$ and two components in the transverse plane $q_{out}$ and $q_{side}$. Where $q_{out}$ is defined parallel, and $q_{side}$ perpendicular to $k_{t}$.
The correlation function is parametrised by a gaussian
\begin{eqnarray}
C_{2}(q)_{th}=1+\lambda\cdot\exp(-R_{o}^{2}q_{o}^{2}-R_{s}^{2}q_{s}^{2}-R_{l}^{2}q_{l}^{2}-2R_{ol}^{2}q_{o}q_{l})
\end{eqnarray}
and the parameters $R_{out}$, $R_{side}$, $R_{long}$, $R_{outlong}$, and $\lambda$ are determined 
by a fit to the measured correlation function. 
The calculations are done in the longitudinal rest frame of the pair (LCMS) where the crossterm 
$R_{ol}$ is supposed to vanish at mid rapidity under the assumption 
of a longitudinally boost invariant expansion \cite{A.3}.
For a chaotic source the parameter $\lambda$ is expectd to be unity and $C_{2}(q)_{th}$ should reach the value 2 
for $q\rightarrow0$. However, due to contaminations of the sample with non-pions and pions from long lived resonances or weak decays, the 
height of the measured correlation function is reduced. To account for 
this effect we weight $C_{2}(q)_{th}$ by a factor $p$ determined by simulations, 
which corresponds to the fraction of correlated pairs in the sample.
Another major effect determining the shape of the measured correlation function is the Coulomb interaction of charged particles.
It is accounted for by a factor $W(q,r_{m})$, so that the final fit-function is given by:
\begin{eqnarray}
C_{2}(q)_{f}= p\cdot(C_{2}(q)_{th}\cdot W(q,r_{m}))+(1-p).
\end{eqnarray} 
The Coulomb-weight $W(q,r_{m})$ depends on the momentum difference $q$ as well as on the mean pair separation $r_{m}$ of the particles in the source. 
It is calculated using a method derived by Sinyukov \etal\cite{A.4} which also provides a 
prescription to derive the value of $r_{m}$ from the extracted source parameters $R_{out}, R_{side},$ and $R_{long}$.
We therefore determine the source parameters as well as the value of $r_{m}$ in an iterative 
fit-procedure. 
The sample was divided into 5 bins in $k_{t}$ in the range 0.0-0.5 GeV/c. 
In longitudinal direction the binsize was chosen to be 0.4 (20, 30, 40 A$\cdot$GeV) and 0.5 (80, 160 A$\cdot$GeV) units of 
pair rapidity.
An example for the projections of the 3-dimensional correlation function 
and the corresponding fits is given in Figure 1.

\begin{figure}[t]
\begin{center}
\epsfig{file=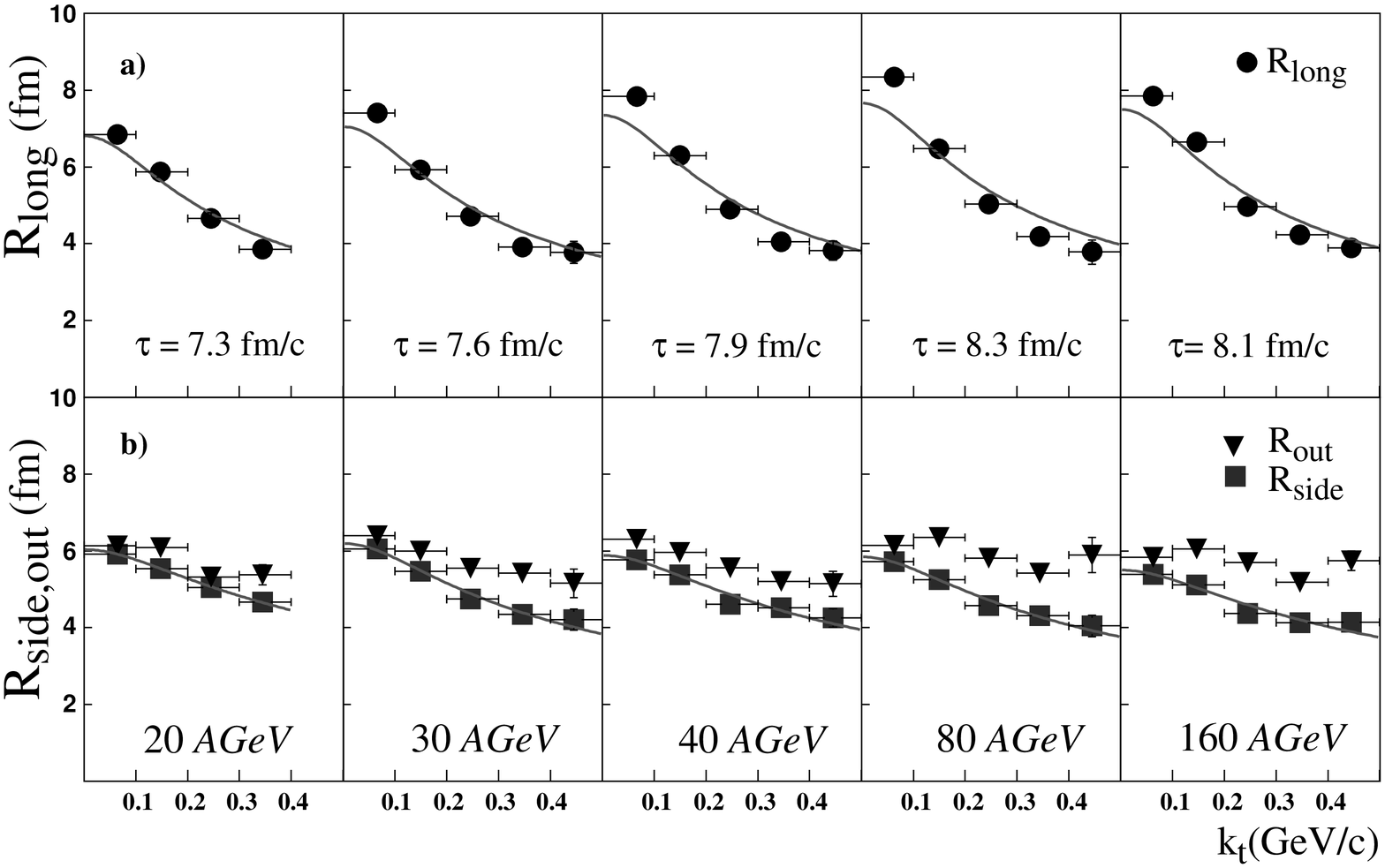,width=13cm}
 
\begin{minipage}{13cm}
\begin{center}

\parbox[t]{10.2cm}{\footnotesize{{\bf Figure 3 :} $k_{t}$-dependence of HBT-radii at midrapidity for central Pb-Pb 
collisions at 20 to 160 A$\cdot$GeV  beam energy as measured by NA49.
The solid lines correspond to fits of eqns. 5 and 6 to $R_{long}$ and $R_{side}$, respectively.}}

\end{center}
\end{minipage}
\end{center}
\end{figure}
\section{Results}
The data presented are not corrected for momentum resolution. The systematic errors of the radii are about 0.5 fm.
Since the Radii are measured in the LCMS 
only a weak rapidity dependence of $R_{out}$, $R_{side}$, and $R_{long}$ is observed 
at all energies. 
However, the finite extension  of the source leads to deviations from the expected  
boost invariant expansion which shows up in an increase of the value of $R_{outlong}$ with increasing 
rapidity as shown in Figure 2 for 30 A$\cdot$GeV beam energy. 
Longitudinal and transverse expansion lead to a reduction of the correlation-lengths with $k_{t}$, which is partly compensated
by a superimposed thermal velocity-field. 
In case of a Bjorken scenario the $k_{t}$-dependence of $R_{long}$ reflects the 
life time of the source \cite{A.5}: 
\begin{eqnarray}
R_{long} & = & \tau_{f}(T_{f}/m_t)^{1/2}\quad,\quad   m_{t}=(m_{\pi}^2+k_{t}^2)^{1/2}.
\end{eqnarray}
Under the assumption of a freezeout temperature $T_{f}$ = 120 MeV a slight increase of 
life time with energy is observed (Figure 3.a).
Assuming a hydrodynamical model with transverse expansion characterized by a linear flow profile with rapidity $\eta_{f}$, the transverse 
geometrical radius $R_{geo}$ of the source can be extracted from the $k_{t}$-dependence of $R_{side}$ \cite{A.6}: 
\begin{eqnarray}
R_{side} & = & R_{geo}/(1+m_{t}\cdot\eta_{f}^{2}/T_{f})^{1/2}. 
\end{eqnarray}
From fits to the data in Figure 3.b we obtain values for the geometrical radius $R_{geo}$ of about 
7-9 fm, but we do not observe a significant energy dependence.
Another important parameter is the emission duration of the source 
which can be determined by \cite{A.7}: 
\begin{eqnarray}
\Delta\tau^{2}& = & \frac{1}{\beta_{t}^{2}}(R_{out}^{2}-R_{side}^{2}) \quad,\quad \beta_{t}\approx\frac{k_{t}}{m_{t}}.
\end{eqnarray}
As seen in the panels of Figure 3.b the difference between $R_{out }$ and $R_{side}$ is positive for all energies 
and the emission duration comes out to be about 3-4 fm/c.\\[0.4cm]
 In summary the NA49-results only show a weak energy dependence of the extracted HBT-radii and of the corresponding source parameters
at SPS energies. This trend persists if we include results from AGS and RHIC experiments as shown in Figure 4. 
We only see a slight rise of $R_{long}$ starting at SPS energies indicating an increase of the life time of the source. $R_{side}$ 
and $R_{out}$ remain approximately constant over the whole energy range.

\begin{figure}[t]
\begin{center}
\epsfig{file=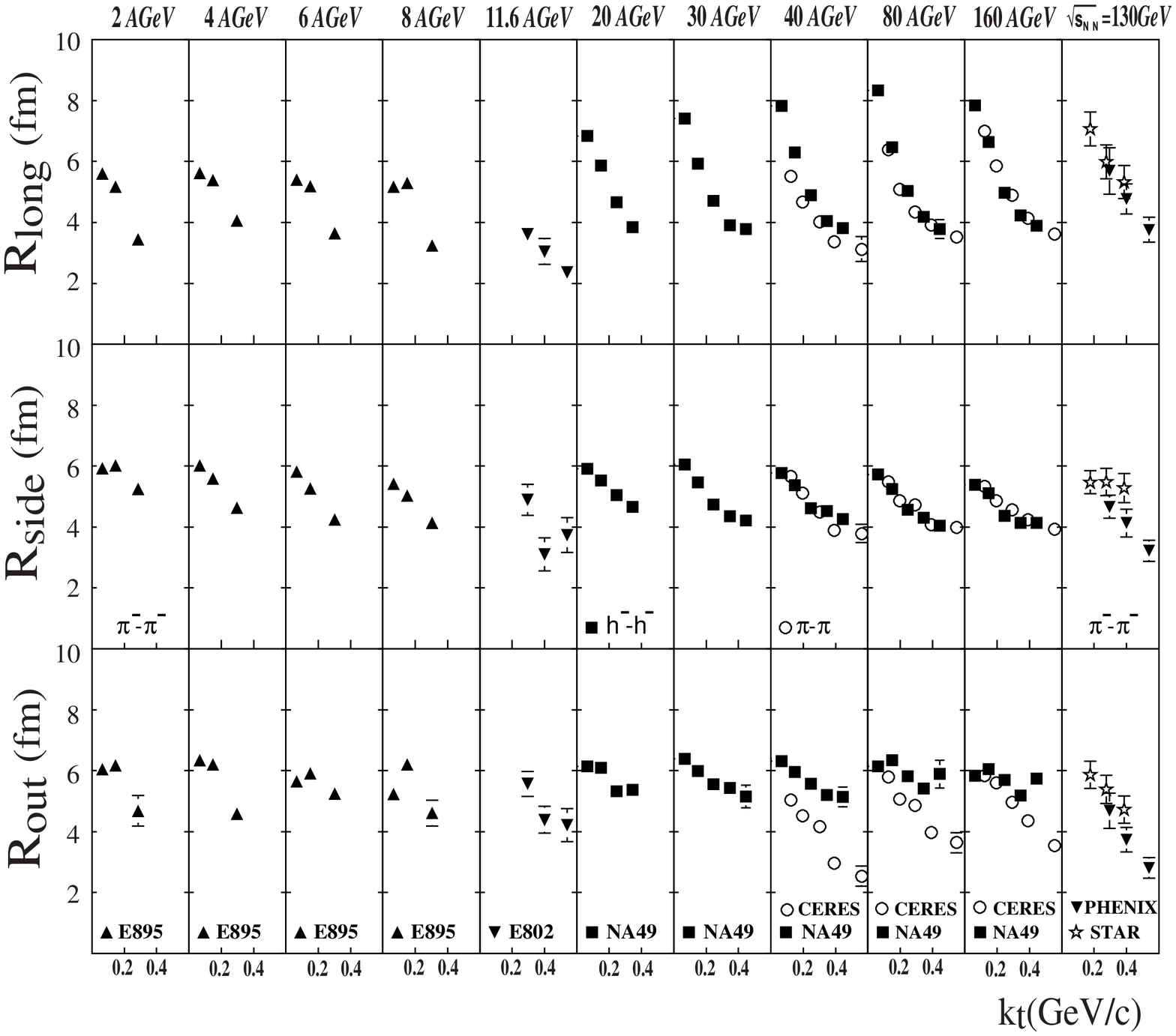,width=13cm}

\begin{minipage}{13cm}
\begin{center}
\parbox[t]{10cm}{\footnotesize{{\bf Figure 4 :} Energy- and $k_{t}$-dependence of the radii $R_{long}$, $R_{side}$, and $R_{out}$ for central Pb+Pb(Au+Au) collisions from AGS to RHIC experiments measured near midrapidity.}}
\end{center}
\end{minipage}

\end{center}				
\end{figure}

\vspace*{0.4cm}

{\bfseries Acknowledgements:} This work was supported by the US Department of Energy
Grant DE-FG03-97ER41020/A000,
the Bundesministerium fur Bildung und Forschung, Germany, 
the Polish State Committee for Scientific Research (2 P03B 130 23, SPB/CERN/P-03/Dz 446/2002-2004, 2 P03B 04123), 
the Hungarian Scientific Research Foundation (T032648, T032293, T043514),
the Hungarian National Science Foundation, OTKA, (F034707),
the Polish-German Foundation, and the Korea Research Foundation Grant (KRF-2003-070-C00015).

\begin{figure}[b]

\end{figure}

\end{document}